# Raman Fingerprints of Phase Transitions and Ferroic Couplings in Van der Waals Material CuCrP$_2$S$_6$

Jing Tang[1], Benjamin J Lawrie[2,3], Mouyang Cheng[4,5,6], Yueh-Chun Wu[2,3], Huan Zhao[2], Dejia Kong[2,7], Ruiqi Lu[1], Ching-Hsiang Yao[1], Zheng Gai[2], An-Ping Li[2], Mingda Li[4,8], Xi Ling[1,9],*

[1]Division of Materials Science and Engineering, Boston University, Boston, Massachusetts 02215, USA
[2]Center for Nanophase Materials Sciences, Oak Ridge National Laboratory, Oak Ridge, Tennessee 37831, USA
[3]Materials Science and Technology Division, Oak Ridge National Laboratory, Oak Ridge, Tennessee 37831, USA
[4]Quantum Measurement Group, MIT, Cambridge, MA 02139, USA
[5]Department of Materials Science and Engineering, Massachusetts Institute of Technology, Cambridge, MA, USA
[6]Center for Computational Science & Engineering, Massachusetts Institute of Technology, Cambridge, MA, USA
[7]Department of Chemistry, University of Virginia, Charlottesville, VA 22903
[8]Department of Nuclear Science and Engineering, MIT, Cambridge, MA 02139, USA
[9]Department of Chemistry, Boston University, Boston, Massachusetts 02215, USA

*Email: xiling@bu.edu

**Abstract**

CuCrP$_2$S$_6$ (CCPS), a type-II multiferroic material, exhibits unique phase transitions involving ferroelectric, antiferroelectric, and antiferromagnetic ordering. In this study, we conduct a comprehensive investigation on the intricate phase transitions and their multiferroic couplings in CCPS across a wide temperature range from 4 to 345 K through Raman spectroscopic measurements down to 5 cm$^{-1}$. We first assign the observed Raman modes with the support of theoretical calculations and angle-resolved polarized Raman measurements. We further present clear signatures of phase transitions from analyses of temperature-dependent Raman spectral parameters. Particularly, two low-frequency soft modes are observed at 36.1 cm$^{-1}$ and 70.5 cm$^{-1}$ below 145 K, indicating the antiferroelectric to quasi-antiferroelectric transition. Moreover, phonon modes hardening is observed when the temperature increases from 4 to 65 K, suggesting negative thermal expansion (NTE) and strong magnetoelastic coupling below 65 K. These findings advance the understanding of vdW multiferroic CCPS, paving the way for future design and engineering of multiferroicity in cutting-edge technologies, such as spintronics and quantum devices.

Multiferroic materials, characterized by the coexistence of ferroelectricity and magnetism, have drawn significant attention due to their potential for applications in spintronics[1], memory devices[2], and multifunctional electronic systems[3]. The coexisting of electric and magnetic order in these materials offers exciting opportunities to manipulate one property via the other, making them highly attractive in the field of electronic devices[4]. Recently, van der Waals (vdW) multiferroic materials emerged for their potential to extend multiferroicity into two-dimensional (2D) frameworks[5,6]. Due to the absence of dangling bonds on the surface at nanometer scale, they could exhibit superior performance in ultra-compact devices[7,8] compared to conventional multiferroic materials.

A notable example for vdW multiferroic materials is monolayer $NiI_2$[9], which exhibits type-II multiferroicity characterized by a proper-screw spin helix coupled with ferroelectric polarization. This intrinsic multiferroic order persists down to the monolayer limit, making $NiI_2$ a promising candidate for 2D multiferroic applications. Another example is the artificial vdW heterostructure composed of bilayer $CrI_3$ and monolayer $Sc_2CO_2$[10]. This engineered system demonstrates both ferromagnetism and ferroelectricity, with a strong magnetoelectric effect that allows tuning of interlayer magnetic coupling via electric polarization switching. Despite these advancements, the number of reported vdW multiferroic materials remains limited. The integration of magnetic elements into ferroelectric vdW materials presents a rare and exciting strategy to achieve multiferroic behavior in 2D systems[11]. This scarcity underscores the need for further discovery of new vdW multiferroics for advanced applications.

$CuCrP_2S_6$ (CCPS), with a crystal structure similar to ferroelectric $CuInP_2S_6$, is a recently synthesized member of the $ABP_2S_6$ family (A and B are transition metals). CCPS exhibits both ferroelectric and magnetic properties because of the substitution of Cr for In, and it is classified as a multiferroic material. Recent study has shown it is a type-II multiferroic vdW material with complex phase transitions[12,13]. In the single layer of CCPS, Cu atoms change their locations in the $[P_2S_6]^{4-}$ framework to introduce ferroelectricity and Cr ions with a net spin of $S = 3/2$ locate in the middle of the layer plane to introduce magnetism (Fig. 1a and b). In its bulk form, CCPS layers are coupled antiferromagnetically (AFM) and $Cr^{3+}$ ions in every layer are ferromagnetically coupled below the Neel temperature ($T_N \approx 32$ K). In addition, multiple other phases are reported in CCPS[13,14], including antiferroelectric (AFE) ordering below $T_{C_1} \approx 145$ K, a quasi-antiferroelectric (QFE) phase above 145 K, a ferroelectric (FE) phase above $T = 190$ K($T_{C2}$), and a paraelectric (PE) phase above 333 K($T_{C3}$). Recent temperature-dependent X-ray diffraction (XRD) measurements down to $T = 70$ K showed distinct changes in interlayer spacing across these phase transitions[15]. Furthermore, atomic resolution transmission electron microscopy at room temperature revealed the alignment of copper atoms to one side of the layers, which is responsible for the ferroelectric ordering in CCPS[14]. Second-harmonic generation (SHG) measurements also corroborated the existence of ferroelectricity at room temperature and above[12]. Importantly, CCPS is classified as a type-II multiferroic, where a strong magnetoelectric (ME) coupling and magnetic-field-induced electric polarization are observe below the Néel temperature. This behavior arises from the spin-dependent $p$–$d$ hybridization due to the slightly off-centered $Cr^{3+}$ ions within sulfur octahedra[13]. Therefore, below $T_N$, CCPS's electric and magnetic orders share a common spin-driven origin, enhancing the coupling between them and suggesting multiferroic behavior even in the monolayer limit.

Despite the above findings, the studies of the vibrational modes and the coupling between ferroic properties and vibrations is still unclear. Understanding this coupling is critical as phonons can mediate interactions between ferroic orders, providing insight into the fundamental mechanisms driving multiferroicity. Raman spectroscopy, a powerful non-invasive technique, is highly sensitive to crystal structural changes and collective excitations[16], making it ideal to study the phase transitions and interactions in the target system. Phase transitions in CCPS refer to changes in its structural and ferroic states as a function of temperature, which are often accompanied by distinctive shifts in vibrational modes or new modes. Previous studies have used Raman spectroscopy to probe phase transitions in various systems. For example, Xi *et al.*, showed that charge density waves exists in monolayer $NbSe_2$ through the measurements of Raman spectra on samples with various number of layers[17]. Lee *et al.*, observed intrinsic antiferromagnetic ordering in $FePSe_3$ through temperature dependent Raman spectroscopy[18]. Song *et al.*, detected helical magnons through circularly polarized Raman spectroscopy in multiferroic $NiI_2$[9]. Recently, the Raman spectra of CCPS and the correlation with several phase transitions were reported.[15,19] Particularly, Susner *et al.* used temperature-dependent Raman spectroscopy to map out the phase transition in the range 70-400 K, with the assist of X-ray diffraction (XRD). Nevertheless, a comprehensive understanding about the Raman spectra such as assignments of Raman peaks, phase-correlated polarization properties, phase-Raman relationship below 70 K, and low-frequency Raman modes remain unclear. In this work, we perform polarized Raman spectroscopy study covering a large frequency range down to 5 cm$^{-1}$ and low temperature down to 4 K to investigate phase transitions and multiferroic properties in CCPS. With the support of density functional theory (DFT) calculation, we assign the experimentally observed phonon modes. Our Raman spectroscopy study further indicates the emergence of multiferroicity, collective excitations and ferroic couplings, shining light to the phase transition processes across a broad temperature range.

CCPS single crystals are synthesized using the chemical vapor transport (CVT) method (see Methods Section) and characterized using X-ray diffraction (XRD) and X-ray photoelectron spectroscopy (XPS) (Fig. 1c and Fig. S1). CCPS flakes are obtained by mechanically exfoliating the bulk crystal using the "Scotch-tape" method. Figure 1d shows a typical atomic force microscopy image of a 4 nm thick flake with a smooth surface characterized by a root mean square (RMS) roughness of 0.23 nm.

CCPS goes through multiple phases as the temperature changes from 4 K to 345 K based on previous literature[13,14]. At the temperature when AFM and AFE phases co-exist (<32 K), CCPS crystallizes in a monoclinic structure belonging to the space group Pc. As shown in Figs. 1a and 1b, the structure consists of layered planes of edge-sharing $(P_2S_6)^{4-}$ octahedra forming a honeycomb-like arrangement, where the $(P_2S_6)^{4-}$ groups consist of two phosphorus atoms connected by a P–P bond, surrounded by six sulfur atoms in a distorted octahedral arrangement. The Cu$^+$ ions and Cr$^{3+}$ ions occupy interstitial sites in these layers and form hexagonal structure in the *ab* plane of each layer. The Cr$^{3+}$ ions have three partially occupied 3*d* orbitals, contributing to intralayer in-plane ferromagnetism and interlayer antiferromagnetism. From the side view, the Cu$^+$ ions occupy one off-center side of layers and Cr$^{3+}$ ions are in the middle of layers. Particularly, the positions of Cu$^+$ ions can shift along the *c* direction as the temperature changes, leading to multiple ferroelectric phases.

In CCPS, the Cu$^+$ ions are sitting in the double-well shape potential straddling the center of the CuS$_6$ octahedron based on previous calculations[20]. The ground state of the monolayer CCPS is AFE and the excitation state is FE with 0.09 eV energy difference per formula unit[21]. It is worth noting that the activation energy from AFE to FE is approximately 0.2 eV per formula unit. This energy is significant for the thermal perturbation (0.2eV / $k_B$ = ~2300 K)[21], so that the thermal perturbation alone is insufficient to excite the AFE ground state, leading to the possibility to observe QFE phase between AFE and FE phases. The competition between thermal effect and order parameters contributes to the complex phase transition of CCPS.

Figure 1e shows the Raman spectra measured on a 50 nm thick CCPS flake at 345 K, 265 K, 165 K, 75 K and 4 K, corresponding to PE, FE, QFE, AFE and multiferroic (AFM and AFE) phase, respectively. At high temperature (i.e. 345 K), the state is governed by the thermal energy and stays in a PE phase where all $Cu^+$ ions are located in the middle of the layers, where the space group of the system belongs to C2/c. The corresponding Raman spectrum shows four peaks at 205, 266, 382 and 601 cm$^{-1}$, respectively. These four peaks have been identified as vibrational modes of transition-metal $Cu^+$ and $Cr^{3+}$ ions (V1 and V2) and symmetric stretching and deformation of $PS_3$ groups (S1 and S2) within the $(P_2S_6)^{4-}$ anions[15]. These four peaks are consistently present and will be analyzed in detail in the following paragraphs. When the temperature decreases to $T$ = 265 K, the double-well shape potential for $Cu^+$ ions in $CuS_6$ octahedron plays a more important role to constrain $Cu^+$ ions at sites with lower internal energy. Thus, $Cu^+$ ions occupy off-center positions on one side of the layers, which results in a FE phase. The Raman spectrum looks similar to that for the PE phase, but the widths of the peaks decrease, due to weaker thermal effects. At $T$ = 165 K, $Cu^+$ ions may be distributed at two possible sites with a certain probability. We use the color fill percentage to represent the probability of the $Cu^+$ ions occupying each site. In this case, the FE order is broken and the system begins to form a quasi-antipolar state, leading to additional peaks at 150, 559, and 617 cm$^{-1}$. When the temperature decreases to $T$ = 75 K, the crystal transits into an AFE phase and the crystal symmetry changes from C2/c to Pc where the unit cell is doubled. Correspondingly, many more Raman signals are observed including pronounced sharp peaks at 280, 300, and 614 cm$^{-1}$. When the temperature reaches $T$ = 4 K, the crystal is in the multiferroic state where AFM and AFE orders co-exist. The Raman spectrum looks similar to that at $T$ = 75 K, and no magnon-Raman modes are observed.

To assign the Raman peaks, we perform angular dependent linear polarized Raman spectroscopy and density functional theory (DFT) calculations on bulk CCPS. To investigate the phase transitions in CCPS in depth, we measure temperature-dependent Raman spectra and investigate the magnetic properties using magnetic property measurement system (MPMS). These analyses reveal a strong magnetoelastic effect resulting from the combination of negative thermal expansion (NTE) and antiferromagnetic-ferromagnetic exchange interaction competition effect in CCPS. Additionally, low-frequency Raman spectroscopy identified two low-frequency soft modes at 36.1 and 70.5 cm$^{-1}$, which clearly correspond to the AFE to FE phase transition. The details are presented below.

Based on group theory, at temperature above $T_{C1}$ (i.e. 145 K), where the crystal exhibits the C2/c symmetry, the primitive unit cell of CCPS contains 20 atoms, resulting in 60 phonon modes, including 3 acoustic modes. At lower temperatures ($T < T_{C1}$), the symmetry changes to Pc, with a doubled unit cell containing 40 atoms, giving rise to 120 phonon modes, including 3 acoustic modes. A normal mode analysis reveals that the lattice vibrations at the Brillouin zone center consist of the following irreducible representations for C2/c point group: $\Gamma = 28Ag + 29Bg$, and for Pc point group: $\Gamma = 58A' + 59A''$. All modes here are Raman active. To assign Raman peaks observed in different phases, angle-resolved polarized Raman spectra are measured at several temperatures, as shown in Figs. S2. The symmetry and Raman tensor components of each peak are analyzed. The intensity ($I$) of a polarized Raman signal (I) is given by:

$$I \propto |\hat{e}_s \cdot R \cdot \hat{e}_i|^2 \ldots\ldots\ldots\ldots(1)$$

Where $\hat{e}_i$ and $\hat{e}_s$ are the normalized incident light polarization vector and scattered light polarization vector, respectively, and $R$ is the Raman tensor of a given mode. For the monoclinic system of CCPS in the AFE phase, the Raman tensors of $A'$ and $A''$ modes can be expressed in the following forms:

$$A' = \begin{pmatrix} \tilde{b} & 0 & \tilde{d} \\ 0 & \tilde{c} & 0 \\ \tilde{d} & 0 & \tilde{a} \end{pmatrix}$$

$$A'' = \begin{pmatrix} 0 & \tilde{f} & 0 \\ \tilde{f} & 0 & \tilde{e} \\ 0 & \tilde{e} & 0 \end{pmatrix}$$

where every tensor element is a complex number with a phase to fine tune the polarization of the peak. Figure 2a shows Raman spectra of CCPS under parallel and cross-polarized configurations at 4 K. 28 $A'$ modes and 16 $A''$ modes are observed out of the predicted 117 optical phonon modes. Their symmetries are assigned by fitting the angular dependent Raman intensity with Eq. 1. The $A'$ and $A''$ modes are labeled with orange and green arrows, respectively. We present the polar plots of the polarized angular dependent Raman intensities of 12 peaks under parallel configuration in Fig. 2b. The two low-frequency modes at 36.1 and 70.5 cm$^{-1}$ are $A'$ modes with relatively weak polarization. Peaks at 86.2, 124.6, 197.5, 202.6 (V1), 266.1 (V2), 301.3, 380.6 (S1), and 599.7 cm$^{-1}$ (S2) are also $A'$ modes, where the peaks at 86.2 and 301.3 cm$^{-1}$ show high degree of polarization. Peaks at 327.7 and 614.4 cm$^{-1}$ are $A''$ modes with clear four-fold symmetry. The degree of polarization (DOP = $\frac{I_{max}-I_{min}}{I_{max}+I_{min}}$) for the four peaks present at all measured temperatures is extracted at six temperatures (i.e. 4, 75, 165, 220, 298, and 345 K) (Fig. 2c). It is clear that the DOPs of V1 and S2 peaks decrease significantly when the temperature increases across $T_{C1}$, implying a structural phase transition from an antipolar phase to a polar or non-polar phase. In this process, as the symmetry of the crystals become higher, the DOPs of Raman peaks become smaller. V2 and S1 peaks may couple weakly with the ferroelectric order, so their DOPs show smaller change compared to V1 and S2 peaks.

To understand the specific vibrational modes of CCPS, DFT calculations are performed to determine the phonon projected density of states (pPDOS) and the vibrational modes at the Γ point. Figure 3 provides a comprehensive overview of these results: Figure 3a presents the pPDOS for the AFE order in CCPS. The contributions from different atomic species are distinct. The heavier atom, Cu, predominantly contribute to the low-frequency modes (from 0 to 150 cm$^{-1}$). Thus, the low-frequency modes are very likely sensitive to the phase transition related to the motion of Cu atoms. Particularly, the calculated modes at 42.0 and 73.7 cm$^{-1}$ are assigned to soft modes and are closely related to the low-frequency peaks observed experimentally at approximately 36.1 and 70.5 cm$^{-1}$, which will be further discussed in subsequent sections. Sulfur (S) atoms contribute significantly to the overall phonon dispersion and are distributed across a wide frequency range because S atoms form direct bonds with Cu, Cr and P. While chromium (Cr) atoms contribute more to mid-frequency modes (from 50 to 360 cm$^{-1}$) and phosphorus (P) atoms mainly contribute to higher-frequency modes (from 500 to 620 cm$^{-1}$). These findings align well with the influence of atomic mass and bonding environment to the vibrational frequency in the crystal structure. The $A'$ and $A''$ modes at the $\Gamma$ point are extracted and compared to the experimental Raman modes, as shown in Fig. 3b. The theoretical and experimental Raman modes exhibit similar distributions, particularly for the modes around 450 cm$^{-1}$, where a strong match is observed. To facilitate this comparison, a linear scaling factor is applied to the theoretical pPDOS and $\Gamma$-point modes. The unscaled results, along with the pPDOS for ferroelectric (FE) CCPS, are provided in Table S2, Table S3 and Figs. S3 in the supplementary material. The experimental modes are assigned by correlating their Raman shifts and symmetries with the theoretical results (Table S2 and Table S3). Figure 3c highlights the vibrational configuration of the four prominent peaks V1, V2, S1 and S2 for AFE order, where the atomic displacements are shown by the green arrows. Modes V1 and V2 calculated at 199.6 and 265.2 cm$^{-1}$ are characterized by significant movements of the metal atoms (Cu and Cr), corresponding to experimental peaks at 202.6 and 266.1 cm$^{-1}$, while modes S1 and S2 calculated at 381.8 and 599.4 cm$^{-1}$ are dominated by the vibrations of $(P_2S_6)^{4-}$ framework, correlating with high-frequency modes observed at 380.6 and 599.7 cm$^{-1}$. These modes reflect the interplay between atomic displacements and structural changes. The theoretical and experimental analyses of pPDOS and vibrational modes provide crucial insights into the vibrational dynamics and phase transitions of CCPS. Further details on temperature-dependent Raman spectra and their implications for phase transitions will be discussed in the subsequent sections.

Figure 4a shows the color plot of the Raman intensity over the temperature range of 4 K to 345 K. Several significant changes in the Raman spectra are observed as the temperature crosses $T_C$. The intensities of all peaks decrease sharply when the temperature increases beyond 145K ($T_{C1}$), which corresponds to the phase transition from AFE to QFE. To illustrate this behavior, Fig. 4b provides insights for the temperature dependent intensity change of peaks at 280, 300 and 640 cm$^{-1}$ and peaks V1, V2, S1 and S2, which appear at 202.6, 266.1, 380.6 and 599.7 cm$^{-1}$, respectively. Peaks V1, V2, S1 and S2 remain visible above the transition temperature of 145 K despite a significant intensity drop, but many other peaks are nearly completely suppressed above the transition temperature of 145 K. We further plot the temperature dependence of the Raman shifts and full width at half maximum (FWHM) of peaks V1, V2, S1 and S2 in Fig. 4c-f. Note that the frequency of a phonon mode without coupling effects is expected to exhibit linear temperature dependence because lattice thermal expansion is generally linearly proportional to the change in temperature.

In contrast, phonon frequencies with a nonlinear temperature dependence suggests the existence of coupling effects. One explanation is that when higher-frequency optical modes scatter into two acoustic phonons with opposite momenta and half of the energy, the phonon frequency change ($\Delta\omega = \omega - \omega_0$) and FWHM ($\Gamma$) are described by the quasi-harmonic model (QHM) proposed by Cowley[22,23]:

$$\Delta\omega(T) = -A\left(\frac{2}{e^{\hbar\omega_0/2k_BT} - 1}\right)$$

$$\Gamma(T) = \Gamma_0\left(1 + \frac{2}{e^{\hbar\omega_0/2k_BT} - 1}\right)$$

where $\Gamma_0$ is the linewidth at zero Kelvin, $\omega_0$ is the zone-center phonon energy at zero Kelvin, $\hbar$ is the reduced Planck constant, $k_B$ is the Boltzmann factor, and T is the absolute temperature. The dashed lines in Figs. 4e-f show the QHM fitting for peak S1 and S2. Among these four persistent peaks, peak S1 approximately follows the QHM, with both its Raman shift and FWHM changing monotonically with temperature. While the frequency of peak S2 deviates from the QHM at higher temperatures, it still changes monotonically with temperature. In the previous calculation, modes S1 and S2 are primarily dominated by the vibrations of the $(P_2S_6)^{4-}$ group, whose structure rarely changes with temperature. The stability of this vibrational group explains why their behaviors roughly align with the QHM. Another two peaks V1 and V2 deviate from QHM behavior significantly and exhibit abrupt changes in Raman shift and FWHM across distinct temperature ranges. We attribute the abrupt changes to the phase transitions in CCPS, where each phase is highlighted in color and the transition temperatures are indicated by black solid lines located at $T_N$ = 30.5, $T_{C1}$ = 145, $T_{C2}$ = 190 and $T_{C3}$ = 333 K. Additionally, the potential phase transitions, marked by red dashed lines at $T$ = 65 and 245 K will be discussed below. In the calculated pPDOS, modes V1 and V2 are characterized by significant movements of the metal atoms (Cu and Cr). The large atomic displacements in these modes explain the pronounced changes in the low-frequency Raman peaks with temperature. This sensitivity can be attributed to the substantial differences in Cu atom positions between the AFE and QFE phases. The symmetry of the lattice and the coupling between metal atoms undergo substantial changes during the AFE to QFE phase transition, resulting in Raman shifts that deviate from the QHM. We also notice that even though peak S1 is aligned well with the QHM generally, slight divergence around the phase transition temperatures is observed.

For example, from 4 K to 65 K, the Raman shift shows a slight hardening with increasing temperature before abruptly softening above 65 K. In Fig. 4c, 4d and 4e, the hardening behaviors are highlighted by orange arrows. This behavior is likely indicative of NTE, which was observed in XRD studies,[15] suggesting the existence of strong magnetoelastic coupling,[24] where the magnetic ordering directly influences lattice vibrations and the phonon modes associated with NTE exhibit behavior opposite to typical thermal expansion trends. In systems with strong spin-lattice interactions, such as spinel LiGaCr$_4$S$_8$, it exhibits NTE between 10.3K and 111 K, as observed through neutron diffraction studies, and magnetic susceptibility measurements show a departure from Curie-Weiss behavior below 110 K[24]. Similar behavior was also observed in the vdW ferromagnet CrBr$_3$, where NET and anomalous phonon hardening occur below its Curie temperature (~37 K), driven by spin–phonon couplings as ferromagnetic order sets in[25]. In CCPS, our zero-field-cooled (ZFC) temperature-dependent magnetization (M-T) measurements results (Fig. S4) are fitted using the Curie-Weiss law ($\chi = \frac{C}{T-T_N}$, $T_N = 30.5K$), which shows a clear departure from the Curie-Weiss law below 65 K, even though the $T_N$ is much lower. This departure suggests that competing AFM and FM exchange interactions emerge. This competition likely drives strong magnetoelastic coupling in CCPS below 65 K, introducing NTE through the interplay of AFM and FM interactions, which modifies the structural response and strongly influence the lattice dynamics. Raman spectroscopy provides a critical foundation to connect magnetism and elasticity to prove that CCPS exhibits strong magnetoelastic coupling. These findings highlight the potential of CCPS as a promising candidate for hosting three ferroic properties (i.e., ferromagnetism, ferroelectricity and ferroelasticity) in a single material. Between 65 K and 145 K, the Raman shifts of peak S1 and S2 approximately follow the QHM, however, the obvious drop behavior of peak V2 is distinct from that in the temperature region below 65 K, indicating this vibrational mode is closely related to antiferromagnetic exchange interaction. Moreover, peak V2 shows pronounced hardening when the temperature is close to 145 K, indicating the FE phase transition. Meanwhile, temperature-dependent Raman shift of peak S2 exhibits distinct slopes in the 145 K–190 K and 65 K–145 K ranges, suggesting that phonon behavior differs in these two phase regimes. This change in slope likely reflects an electrostrictive coupling, where the lattice dynamics are influenced by the onset or strengthening of ferroelectric order. In particular, the S2 mode involving vibrations within the (P$_2$S$_6$)$^{4-}$ framework, is sensitive to polar displacements[26]. From 190 to 333 K, peak V2 begins to redshift and peak S2 exhibited another distinct shifting rate.

The difference above 333 K is difficult to conclude from our temperature-dependent Raman measurements due to the limited number of data points. Except for the above-mentioned transition temperatures, our results show 245 K is another critical temperature worth of attention. Previous study on temperature-dependent XRD measurements reveal a sharp change in layer spacing near this temperature[15], and peaks S1 and S2 exhibit noticeable shifts around 245 K that are highlighted by green arrows (Fig. 4e and 4f). This behavior may be indicative of a second-order transition. For instance, the potential energy landscape for Cu ions might feature multiple wells, with the Cu ions migrating between these sites during the transition.

In the temperature dependent Raman experiment, two low-frequency soft Raman modes at 36.1 and 70.5 cm$^{-1}$ are of particular interest. Phonon softening is an important indicator of strong coupling between electronic and vibrational degrees of freedom in materials. Low-frequency soft modes are often associated with phase transitions. For instance, in SrTiO$_3$, two low-frequency soft modes at 15 cm$^{-1}$ and 48 cm$^{-1}$ soften to zero frequency as the temperature approaches the transition temperature ($T_C = 110\ K$)[27]. In CCPS, a similar phenomenon is observed in the low-frequency region of the Raman spectrum. However, instead of softening to zero frequency, the modes soften to a finite frequency as $T \rightarrow T_C$. Figure. 5a presents the color plot of the temperature-dependent low-frequency Raman spectra, where peaks at 36.0 cm$^{-1}$ (SM$_1$) and 70.5 cm$^{-1}$ (SM$_2$) fitted with Lorentzian peaks (Figure S5) exhibit distinct temperature-dependent behavior compared to other peaks at 85 and 88 cm$^{-1}$. The trends of SM$_1$ and SM$_2$ are marked by blue and pink dashed lines, respectively. As the temperature increases, these peaks redshift rapidly and disappear near $T_C = 145$ K. From the angular polarized Raman spectra measurements, peak SM$_1$ and SM$_2$ are identified as $A'$ mode. Our DFT calculations presented in previous section show that the low-frequency modes are dominated by Cu atoms' vibrations. We interpret these two soft modes as reciprocal lattice boundary phonon modes that fold to the Brillouin zone center when a lower-symmetry lattice forms. Concurrently, the unit cell size doubles. The temperature-dependent frequency of the soft mode can be fitted as[27]: $\omega = A\ (T_C - T)^n$ with $T_C = 145\ K$, where $A$ is a constant (Fig. 5b). For peak SM$_1$, $n = 0.46 \pm 0.04$, and for peak SM$_2$, $n = 0.38 \pm 0.04$. The $n$ value indicates how rapidly the soft mode frequency changes as the system approaches the critical temperature, which is determined by the vibrational properties of the specific soft mode.

In conclusion, we studied the vibrational properties and ferroelectric phase transitions in the type-II multiferroic vdW material CCPS via Raman spectroscopy. We assigned the experimentally observed Raman peaks through angle-resolved polarized Raman spectroscopy measurements and DFT calculation. This information provides a solid foundation for future use of Raman spectroscopy as a convenient tool to study the multiferroic CCPS. Moreover, the temperature-dependent Raman spectra provide fingerprints about intricate phase transitions and ferroic couplings in CCPS. Several important phase transitions at $T$ = 65, 145, 190 and 245 K are discussed to understand the origin of their influence on the phonon modes and ferroic mechanism of CCPS.

**Methods**

Sample preparation: CCPS single crystals were synthesized using a chemical vapor transport (CVT) method. Stoichiometric amounts of high-purity elemental copper (Cu), chromium (Cr), phosphorus (P), and sulfur (S) were thoroughly mixed and sealed in a quartz ampoule under vacuum ($1\times10^{-4}$ torr). Iodine ($I_2$) was used as the transport agent at a concentration of approximately 20~30 mg/g. The ampoule was then placed in a two-zone furnace, with the source zone heated to a temperature of 800 °C and the growth zone maintained at 750 °C. Over a period of 12 days, the temperature gradient facilitated the transport of the material from the hotter to the cooler zone, leading to the growth of high-quality CCPS crystals. CCPS flakes were obtained by mechanically exfoliate the bulk crystal using the "Scotch tape" method[28].

XRD and XPS characterizations: XRD results were collected at room temperature for 1 hour with a Bruker D8 Discover in parallel beam geometry, with a sealed tube Cu X-ray source operating at 1600 watts, Goebel mirror monochrometer, a 0.6 mm primary divergence slit, 2.5 degree primary and secondary axial Soller slits and a LynxEye 1D detector. X-ray Photoelectron Spectroscopy (XPS) was performed using the PHI Genesis system, a state-of-the-art instrument designed for precise surface analysis, utilizing monochromatic Al Kα radiation (1486.6 eV) to probe the elemental composition and chemical states of the CCPS. The PHI Genesis features high sensitivity and energy resolution, enabling reliable detection of core-level peaks. Data acquisition and processing were performed under ultra-high vacuum conditions ($3\times10^{-7}$ Pa), ensuring minimal contamination and accurate spectral analysis.

Raman spectroscopy characterizations: Polarization-resolved Raman spectra were collected over the temperature range of 4 K to 345 K using a Montana Instruments closed-cycle cryostat with out-of-plane excitation and backscattering geometry. A 532.03 nm continuous-wave laser was used as the excitation source, delivering approximately 2 mW of power at the sample. The Raman spectra were measured using a Princeton Instruments Isoplane SCT-320 spectrograph equipped with a Pixis 400BR Excelon camera and a 2400 lines/mm grating. Low-energy Stokes Raman modes were accessed using a set of three Optigrate volume Bragg gratings. For polarization control, achromatic half-wave plates were mounted on piezoelectric rotators. Each spectrum typically required an acquisition time of approximately 120 seconds.

Magnetic measurements: Magnetic measurements were performed using a Quantum Design Magnetic Property Measurement System (MPMS) at the Center for Nanophase Materials Science (CNMS) at Oak Ridge National Laboratory. The zero-field cooled (ZFC) measurements were carried out between the temperature range of 2-400 K either in plane or out of plane of the flake. The in plane and out of plane ZFC magnetizations are shown in Fig. S4. A magnetic transition shows at 31.5 K, with a sharp cusp at the transition temperature. No other magnetic transitions are observed all the way up to 400 K.

DFT calculations: Density Functional Theory (DFT) calculations were conducted for the anti-ferroelectric (AFE) phase of $CuCrP_2S_6$ using the Vienna Ab initio Simulation Package (VASP)[29]. The projector-augmented wave (PAW) method[30,31] was employed, with exchange-correlation effects described by the Perdew-Burke-Ernzerhof (PBE) formulation of the generalized gradient approximation (GGA)[32]. The plane-wave cutoff energy was set at 400 eV to ensure sufficient convergence. A calibrated Hubbard U correction of 3.7 eV, as used by the Materials Project[33], was applied to better account for the on-site Coulomb interactions in the Cr d orbitals. Calculations were performed with a 5 × 3 × 3 k-point mesh centered at the Gamma point, to sample the Brillouin zone. Geometry optimization was carried out with a convergence criterion for the forces set to 0.01 eV / Å under the symmetric constraint of the Pc space group. The vibrational modes and irreducible representations were calculated by combining VASP with the Phonopy package[34] with a 2 × 1 × 1 supercell. Additionally, the Bilbao Crystallographic Server[35] was consulted to identify the Raman-active modes.

**Supporting information**

X-ray photoelectron spectroscopy (XPS) of CVT synthesized CuCrP$_2$S$_6$ (CCPS), Full-Angle Polarized Raman Spectra on CCPS: 2D Color Plots of Z(XX)Z and -Z(XY)Z Configuration, Complex Raman tensors of CCPS, Comparison between calculated and experimental Raman shift data, DFT calculated projected phonon density of states (pPDOS) for CCPS with antiferroelectric (AFE) and paraelectric (PE) order, Zero-field-cooled (ZFC) temperature-dependent magnetization (M-T) of CCPS, Low-Frequency Raman Spectrum Analysis.

**Acknowledgments**

This material is based upon work supported by the National Science Foundation (NSF) under Grant 1945364. Work done by X.L. was also supported by the U.S. Department of Energy (DOE), Office of Science, Basic Energy Science (BES) under Award DE-SC0021064 and NSF under Grant No. 2216008. The Raman spectroscopies and magnetic properties measurement were supported by the Center for Nanophase Materials Sciences (CNMS), which is a US Department of Energy Office of Science User Facility at Oak Ridge National Laboratory with additional support for cryogenic Raman spectroscopies provided by the U. S. Department of Energy, Office of Science, Basic Energy Sciences, Materials Sciences and Engineering Division. The XRD measurement was supported by NSF under the Award No. 1337471 and the BU MSE Core Research Facility. The computational work done at MIT is supported by DOE BES Award DE-SC0021940 and NSF Award DMR-2118448. J.T. and X.L. also thank Riichiro Saito and Nguyen Tuan Hung at the Tohoku University in Japan for the discussion on DFT calculations on the Raman spectra of CCPS, and Anna Swan from Boston University for her valuable discussions on the Raman spectra analysis.

**Author contribution**

J.T. and X.L. conceived the idea and experimental plan. J.T, B.J.L., Y.W. and H.Z. performed the spectroscopy measurements. D.K., Z.G. and A.P.L. performed the magnetic properties measurements. R.L. performed the XPS measurement. C.H.Y. performed the XRD measurement. M.C. and M.L. performed DFT calculations. J.T. and X.L. wrote the manuscript with inputs from all other authors.

**Notes**

The authors declare no competing interests.

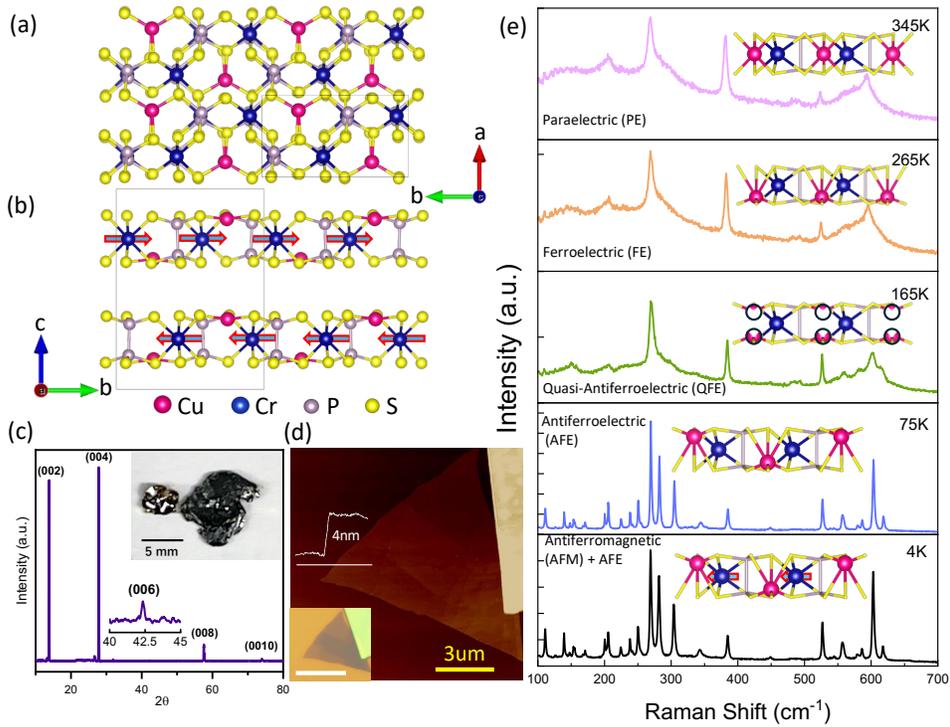

**Figure 1.** (a) Top view (ab-plane) of the CCPS crystal structure at $T = 4$ K. (b) Side view (bc-plane) of the CCPS crystal structure at $T = 4$ K. (c) XRD pattern of CCPS measured at room temperature with an optical image of the crystals in the inset. (d) Atomic force microscopy image of an exfoliated thin layer of CCPS. Inset: optical image of the CCPS flake. The scale bar is 10 μm. (e) Raman spectra of CCPS at various temperatures: $T = 4$ K, 75 K, 165 K, 265 K, and 345 K. The inner crystal structures illustrate the ferroic order in CCPS, with the specific order labeled at the upper left corner of each spectrum.

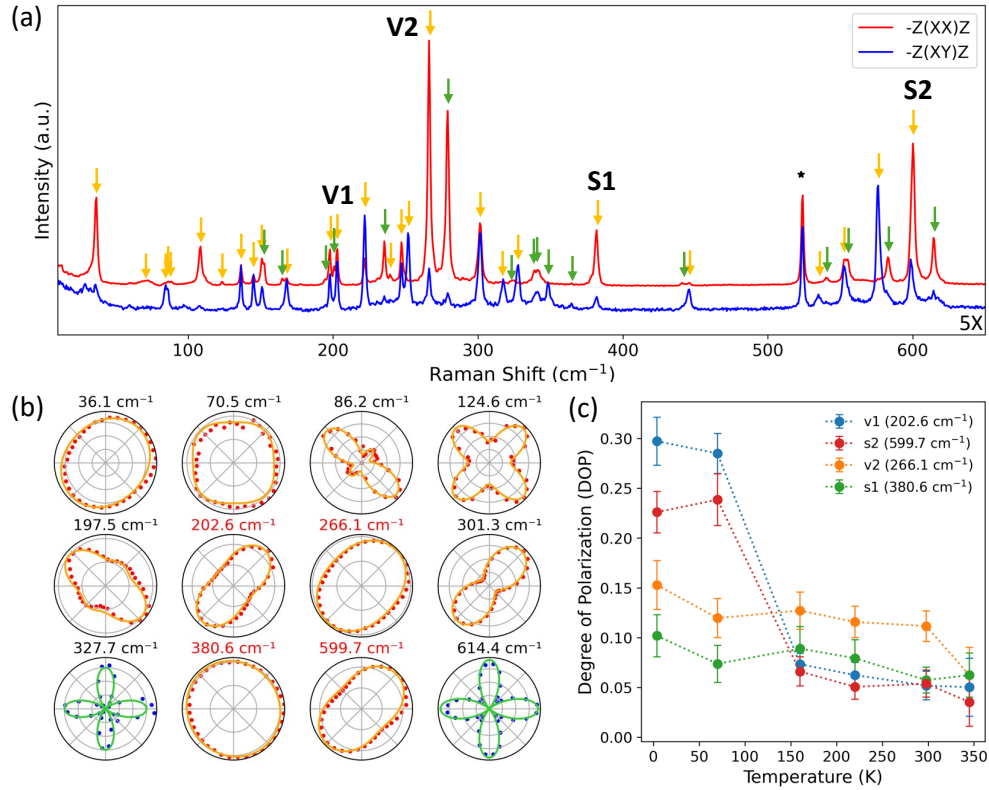

**Figure 2.** (a) Raman spectra of CCPS at 4 K in parallel (red solid curve) and cross (blue solid curve) configurations. The vertical orange arrows and green arrows indicate the $A'$ and $A''$ modes, respectively. (b) Polar plots showing the angular dependence of the intensities for selected peaks, with the corresponding wavenumbers of the peaks labeled above each plot. The wavenumbers of the four persistent peaks (V1, V2, S1, and S2) are highlighted in red. Experimental intensities for the -Z(XX)Z configuration are represented by red dots. The fitted curves for $A'$ and $A''$ modes are represented by orange and green solid lines, respectively. (c) Temperature-dependence of the DOP for peaks V1, V2, S1, and S2.

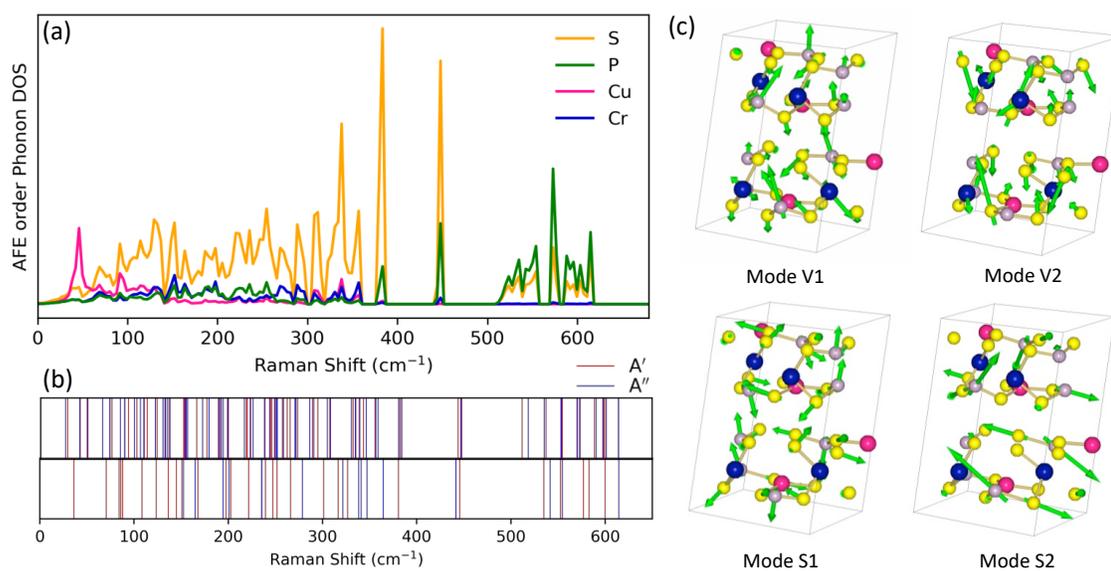

**Figure 3.** (a) Phonon projected density (pPDOS) of states for AFE CCPS calculated using DFT. (b) Comparison of theoretical (top) and experimental (bottom) phonon modes, with vertical red lines indicating $A'$ modes and vertical blue lines indicating $A''$ modes. (c) Computed vibrational modes corresponding to peaks V1, V2, S1, and S2. The green arrows represent the atomic displacements, with arrow lengths indicating the magnitude of the movements.

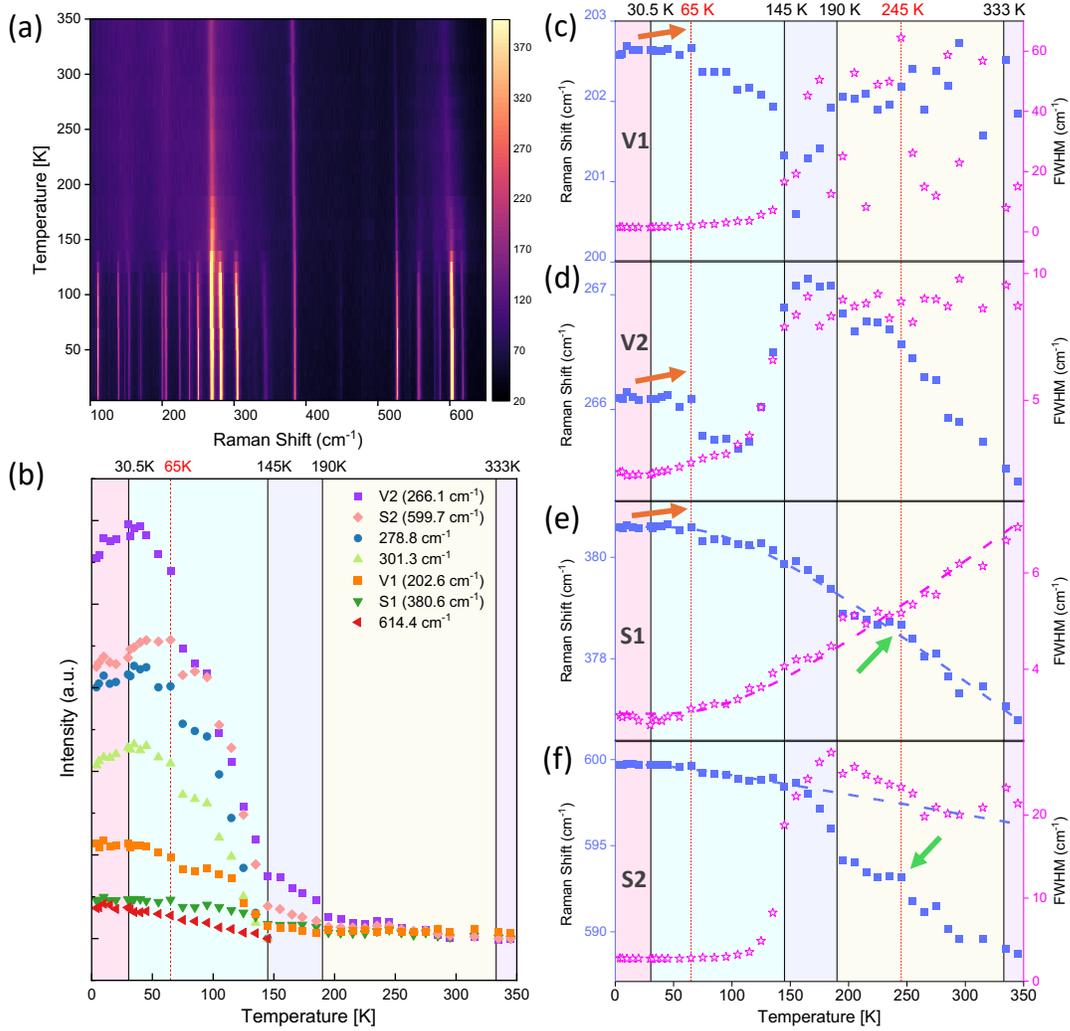

**Figure 4.** (a) Temperature-dependent high-frequency Raman 2D color plot of CCPS from 4 to 345 K. (b) Temperature-dependence of Raman intensities of peaks at 280, 300, 614 cm$^{-1}$ and peak S1, S2, V1 and V2. (c–f) Temperature-dependent Raman shift and FWHM of peaks V1, V2, S1, and S2, respectively. The plots are color-coated to reflect different phases. Orange and green arrows indicate subtle changes in the data.

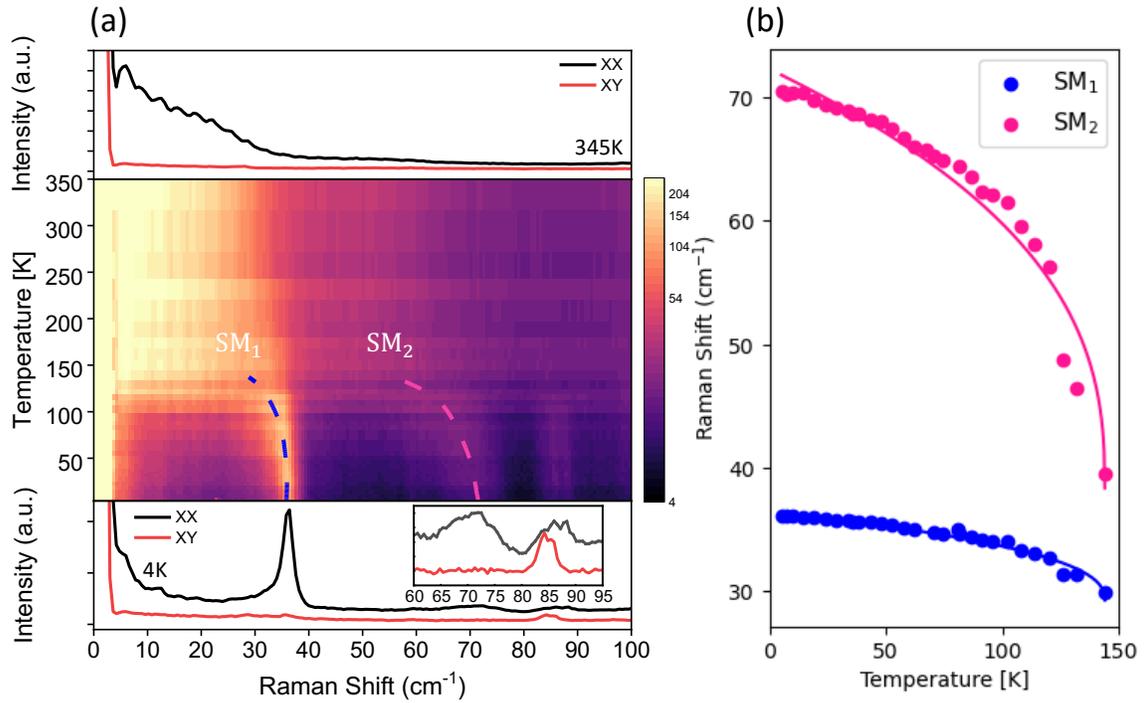

**Figure 5.** (a) Temperature-dependent low-frequency soft Raman modes of CCPS in 2D color plot from 4 K to 345 K. The two soft modes are highlighted by blue and pink dashed lines. The bottom inset shows the low-frequency Raman spectra in parallel and cross-polarized configurations at $T = 4$ K, while the top inset shows the corresponding spectra at $T = 345$ K. (b) Temperature-dependence of the Raman shifts of the two soft modes, with the fitted curves represented by pink and blue solid lines.